\documentclass[11pt]{article}
\usepackage[textwidth=15.2cm,textheight=22cm]{geometry}
\usepackage{amsmath,amssymb}
\usepackage{latexsym}
\usepackage{multicol}
\usepackage{graphicx}
\usepackage{bm}
\allowdisplaybreaks[1]

\newcommand{\del}{\partial}
\newcommand{\be}{\begin{equation}}
\newcommand{\ee}{\end{equation}}
\newcommand{\ba}{\begin{eqnarray}}
\newcommand{\ea}{\end{eqnarray}}
\newcommand{\bdm}{\begin{displaymath}}
\newcommand{\edm}{\end{displaymath}}

\def\ba{\bar A}

\def\beq{\begin{equation}}
\def\eeq{\end{equation}}

\newcommand{\nn}{\nonumber}

\newcommand{\ndt}{\noindent}

\def\bea{\begin{eqnarray}}
\def\eea{\end{eqnarray}}
\def\beas{\begin{eqnarray*}}
\def\eeas{\end{eqnarray*}}
\def\sla{\raise.15ex\hbox{$/$}\kern-.57em}

\def\parm{{\partial}_{-}}

\def\spa#1.#2{\left\langle#1\,#2\right\rangle}
\def\spb#1.#2{\left[#1\,#2\right]}

\begin{document}

\begin{titlepage}
\begin{flushright}    
{\small $\,$}
\end{flushright}
\vskip 1cm
\centerline{\Large{\bf{Factorization of cubic vertices involving}}}
\vskip 0.5cm
\centerline{\Large{\bf{three different higher spin fields}}}
\vskip 1.5cm
\centerline{Y. S. Akshay and Sudarshan Ananth}
\vskip .5cm
\centerline{\it {Indian Institute of Science Education and Research}}
\centerline{\it {Pune 411008, India}}
\vskip 1.5cm
\centerline{\bf {Abstract}}
\vskip .5cm
We derive a class of cubic interaction vertices for three higher spin fields, with integer spins $\lambda_1$, $\lambda_2$, $\lambda_3$, by closing commutators of the Poincar\'e algebra in four-dimensional flat spacetime. We find that these vertices exhibit an interesting factorization property which allows us to identify off-shell perturbative relations between them.
\vfill
\end{titlepage}

\section{Introduction}
\ndt Interactions of massless particles are, in general, very highly constrained. In flat spacetime backgrounds, there exist consistent cubic interaction vertices describing massless higher spin fields~\cite{s3v, BBB} and their couplings to gravity~\cite{RRM1}. At the level of the equations of motion, there has been considerable progress in our understanding of higher spin theories in both flat and anti-de Sitter spacetimes. Fully interacting, non-linear equations of motion describing higher spin fields are known~\cite{MV}. However, a consistent description of higher spin fields ($\lambda>2$), at the level of the action, remains elusive. Interesting attempts to address this problem using a Fock space approach and the Poincar\'e algebra include~\cite{BBB, RRM1}. In this paper, we adopt a more direct method, conducive to our aim of identifying factorization properties and establishing perturbative ties in the space of higher spin theories.
\vskip 0.3cm
%new
\ndt In this paper, we work with the Poincar\'e generators for (3+1) dimensional flat spacetime in light-cone gauge. We write down a general ansatz for the cubic interaction vertex in a theory describing three different higher spin fields. Demanding closure of the Poincar\'e algebra yields a class of generic higher spin cubic interaction vertices which we rewrite in momentum space, using spinor helicity notation. 
\vskip 0.3cm
\ndt In this spinor helicity language, interesting structures and relations are manifest. Specifically, the cubic vertex in higher spin theories may be obtained by simply multiplying the corresponding cubic vertices involving lower spin fields. Given two theories, one involving spins $(\lambda_1, \lambda_2, \lambda_3)$ and the other $(\lambda_1', \lambda_2', \lambda_3')$, one may obtain by the direct product of their cubic vertices, the cubic interaction term for a theory involving spins $(\lambda_1+\lambda'_1, \lambda_2+\lambda'_2, \lambda_3+\lambda'_3)$. A converse of this property is the factorization of the higher spin vertex into the corresponding lower spin ones.
\vskip 0.3cm
\ndt One obvious consequence of these properties is that the cubic vertex describing fields of spins $(n\,\lambda_1, n\,\lambda_2, n\,\lambda_3)$ is the $n^{th}$ power of the $(\lambda_1, \lambda_2, \lambda_3)$ cubic vertex. This is reminiscent of the KLT relations~\cite{KLT} and even more so of their off-shell extensions~\cite{AT}. An interesting question is whether such relations extend to quartic and higher order vertices in higher spin theories. Further study along the lines described here, requires the extension of the Poincar\'e generators and the calculations presented here to order $\alpha^2$. There is however, reason to believe that such an attempt might run into difficulties. The no-go theorems for higher spin theories~\cite{BBS} do not allow consistent interacting theories involving particles of spin greater than two. On the other hand, many of these results were derived assuming that locality and Lorentz invariance were manifest in the theory under consideration. Neither of these properties is manifest in light-cone gauge making it an ideal choice for the study of higher spin fields, a point we return to at the end of this paper. The cubic interaction vertices we derive here are not a comprehensive listing  of all possible vertices~\footnote{For such listings, see for example~\cite{RRM2}.} since our aim here is not to be encyclopaedic but to establish both factorization properties and perturbative links in higher spin theories. 
\vskip 0.3cm

%\subsection{Summary}
%
%\ndt We write down the Poincar\'e generators, for four dimensional flat spacetime, in light-cone gauge. We start with a general Ansatz for the interaction Hamiltonian describing three different higher spin fields. Closure of the Poincar\'e algebra will impose restrictions on this Ansatz. These restrictions will yield a class of generic higher spin cubic vertices. In momentum space, we show that these vertices exhibit factorization properties and perturbative ties, reminiscent of the KLT relations~\cite{KLT} and their off-shell extensions~\cite{AT}.

\section{Poincar\'e generators}

\ndt We define light-cone co-ordinates in $(-,+,+,+)$ Minkowski space-time by
\begin{eqnarray}
x^{\pm}=\frac{x^{0}\pm x^{3}}{\sqrt{2}} \;,\qquad
x = \frac{x^{1}+ix^{2}}{\sqrt{2}} \;,\qquad\bar{x}= \frac{x^{1}-ix^{2}}{\sqrt{2}}\ .
\end{eqnarray}
The corresponding derivatives being $\partial_{\pm}\,,\,\,\bar{\partial}$ and $\partial$. In four spacetime dimensions, all massless fields have two physical degrees of freedom for which we use the notation $\phi$ and $\bar{\phi}$. We choose the field $\phi$ to have helicity $\lambda$ while the field $\bar\phi$ has helicity $-\lambda$. The generators of the Poincar\'{e} algebra, in light-cone coordinates are
\bea
p^{-}=i\frac{\partial\bar{\partial}}{\partial_{-}}=-p_+ \qquad p^+=-i\partial^{+}=-p_- \qquad  \bar{p}=-i\bar{\partial}\qquad p=-i\partial\ ,
\eea
\begin{eqnarray}
j\!\!\!\!\!\!&= i(x\bar{\partial}-\bar{x}\partial - \lambda) \;,\qquad  &j^{+} = (x^+ \partial-x\partial^+)\ , \nn\\
 j^{+-}\!\!\!\!&=(x^{+}\frac{\partial\bar{\partial}}{\partial^+}-x^-\partial^+ ) \;,\qquad & j^-=(x^{-}\partial-x\frac{\partial\bar{\partial}}{\partial^+}+\lambda \frac{\partial}{\partial^+}) \ ,
\end{eqnarray} 
and their complex conjugates. $\frac{1}{\parm}$ is defined using the prescription in~\cite{SM}. \\
Using the free equations of motion $\partial_{+}=\frac{\partial\bar{\partial}}{\partial_{-}}$ which is modified in the interacting theory.
The Hamiltonian for the free field theory is
\begin{equation}
H\equiv\int d^{3}x\,\mathcal{H}=-\int d^3x\,\bar\phi\,\partial\bar\partial\,\phi\  ,
\end{equation}
where the second equality only holds for the free theory. This is rewritten as
\begin{equation}
\label{hamilx}
H\equiv\int d^{3}x\,\mathcal{H}=\int d^{3}x\,\partial_{-}\bar{\phi}\,\delta_{\mathcal{H}}\phi
\ ,
\end{equation}
in terms of the time translation operator
\begin{eqnarray}
\delta_{\mathcal{H}}\phi\equiv\partial_{+}\phi=\lbrace \phi,\mathcal{H}\rbrace\ ,
\end{eqnarray}
where $\lbrace ,\rbrace$ denotes the Poisson bracket. When interactions are switched on, $\delta_{\mathcal H}$ picks up corrections, order by order in the coupling constant $\alpha$. The other generators that pick up corrections are
\begin{eqnarray}
\delta_{j^{+-}}\phi \!\!\!\!\!\!\!\!\!&&= \delta_{j^{+-}}^{0}\phi -ix^{+}\delta^{\alpha}_{\mathcal{H}}\phi + O(\alpha^{2})\ ,\nn \\
\delta_{j^{-}}\phi \!\!\!\!\!\!\!\!\!&&= \delta_{j^{-}}^{0}\phi +ix\delta^{\alpha}_{\mathcal{H}}\phi +\delta^{\alpha}_{s}\phi + O(\alpha^{2})\ , \nn \\
\delta_{\bar{j}^{-}}\phi \!\!\!\!\!\!\!\!\!&&= \delta_{\bar{j}^{-}}^{0}\phi +i\bar{x}\delta^{\alpha}_{\mathcal{H}}\phi +\delta^{\alpha}_{\bar{s}}\phi + O(\alpha^{2}).
\end{eqnarray}
Here, $\delta_{s}^{\alpha}$ and $\delta_{\bar{s}}^{\alpha}$ represent spin transformations. We assume these to be  of the form $\bar{\phi}\phi$ as this form agrees with the known transformations. At cubic order, these do not mix with any of the other terms and are therefore not relevant to the calculations in this paper.

\section{Deriving cubic interaction vertices}
\vskip 0.2cm
\ndt We focus on the following three structures for cubic interaction vertices at order $\alpha$.

\bea
\delta^{\alpha}_{\mathcal{H}}\phi_1\sim\phi_2\phi_3\ ; \quad \delta^{\alpha}_{\mathcal{H}}\phi_2\sim\phi_1\phi_3\ ;\quad \delta^{\alpha}_{\mathcal{H}}\phi_3\sim\phi_1\phi_2\ ,
\eea
where the fields $\phi_1$, $\phi_2$ and $\phi_3$ have integer spins $\lambda_1$, $\lambda_2$ and $\lambda_3$ respectively. The first of these structures, at the level of the action would correspond to terms of the form
\bea
{\mbox {\it {S}}}\sim\int \,d^4x\,\;\bar\phi_1\,\phi_2\,\phi_3\ +c.c.\:.
\eea
We enhance this basic form with derivatives to arrive at the Ansatz 
\begin{equation}
\label{integers}
\delta^{\alpha}_{\mathcal{H}}\phi_1=\alpha\,A\,\partial^{+\mu}\left[\bar\partial^a\partial^{+\rho}\phi_2\bar\partial^b\partial^{+\sigma}\phi_3\right]\ +c.c. ,
\end{equation}
where $\mu ,\rho ,\sigma ,a,b$ are integers and $A$ is a numerical factor that could depend on the variables and spins. Note that terms of the form $\bar{\phi}_2\bar{\phi}_3$ in $\delta_H^\alpha \phi_1$ are independent of terms of the form $\phi_2\phi_3$ and hence will not `talk' to one another. The commutators
 \begin{eqnarray}
&&\left[\delta_{j},\delta_{\mathcal{H}}^{\alpha}\right]\phi_1 =0\ , \nn \\
&&\left[\delta_{j^{+-}},\delta_{\mathcal{H}}^{\alpha}\right]\phi_1 =-\delta_{\mathcal{H}}\phi_1\ ,
\end{eqnarray}
impose the following conditions on our Ansatz
\begin{eqnarray}
\label{conds1}
\nn a+b\!\!\!\!\!\!&&=\lambda_2+\lambda_3-\lambda_1  \\
\mu +\rho +\sigma\!\!\!\!\!\!&&=-1 \ .
\end{eqnarray} 
Since $a,b>0$, the first of these conditions{\footnote{Note that this reduces to the condition in~\cite{BBB} if we set $\lambda_1=\lambda_2=\lambda_3$.} implies that the vertex cannot exist unless $\lambda_2+\lambda_3>\lambda_1$. Now, let $\lambda \equiv \lambda_2 +\lambda_3 -\lambda_1$ so the first equation of (\ref{conds1}) reads $a+b = \lambda$. There are precisely $(\lambda + 1)$ possible values for a pair $(a,b)$. We now rewrite our Ansatz in (\ref{integers}) as a sum of these $(\lambda + 1)$ terms
\be
\label{ansatz}
\delta_H^\alpha \phi_1=\alpha\sum_{n=0}^{\lambda}A_n\,\partial^{+\mu_n}\left[\bar\partial^n\partial^{+\rho_n}\phi_2\,\bar\partial^{(\lambda -n)}\partial^{+\sigma_n}\phi_3\right]\ +c.c. \: .
\ee
\\
\ndt The next set of commutators are 
\bea
\left[\delta_{\bar{j}^{-}}, \delta_{H}\right]^{\alpha}\phi_{1} =0 \qquad \left[\delta_{j^{+}}, \delta_{H}\right]^{\alpha}\phi_{1} = 0\: ,
\eea
and yield the following conditions
\bea
\label{conds3}
\nn \sum_{n=0}^{\lambda}\!\!\!\!\!\!&& A_n\,{\biggl \{}\;(\mu_n +1-\lambda_1)\del^{+(\mu_n -1)}\bar{\del}(\bar{\del}^{n}\del^{+\rho_n}\phi_2\bar{\del}^{(\lambda -n)}\del^{+\sigma_n}\phi_3)\\
&&+(\rho_n +\lambda_2)\del^{+\mu_n} (\bar{\del}^{(n+1)}\del^{+(\rho_n -1)}\phi_2\bar{\del}^{(\lambda -n)}\del^{+\sigma_n}\phi_3)\\	
\nn &&+(\sigma_n +\lambda_3)\del^{+\mu_n} (\bar{\del}^n\del^{+\rho_n}\phi_2\bar{\del}^{(\lambda-n+1)}\del^{+(\sigma_n -1)}\phi_3){\biggr \}}\ =0\ ,\\
\nn\sum_{n=0}^{\lambda}\!\!\!\!\!\!&& A_n{\biggl \{}\;n\,\del^{+\mu_n}(\bar{\del}^{(n-1)}\del^{+(\rho_n +1)}\phi_{2}\bar{\del}^{(\lambda -n)}\del^{+\sigma_n}\phi_{3})\\&&+(\lambda -n)\del^{+\mu_n}(\bar{\del}^{n}\del^{+\rho_n}\phi_{2}\bar{\del}^{(\lambda -n-1)}\del^{+(\sigma_n +1)}\phi_{3}){\biggr \}}=0\ .
\eea
\\
These conditions are satisfied if the coefficients obey the following recursion relations.
\be
\label{A}
\nn A_{n+1}=-\frac{(\lambda - n)}{(n+1)}A_n \,\,=\,\, (-1)^{(n+1)} \binom{\lambda}{n+1}A_0\ ,
\ee
\be
\label{rho}
\rho_{n+1}=\rho_n - 1 \quad;\quad \sigma_{n+1}=\sigma_n +1 \quad ;\quad \mu_{n+1}=\mu_n \: ,
\ee
with the last condition showing that $\mu_n$ is independent of $n$. The following ``boundary" conditions are also necessary.
\be
\label{boundary}
\rho_{{\,}_{n=\lambda}}=-\lambda_2 \qquad \sigma_{{\,}_{n=0}}= -\lambda_3.
\ee
\\ 
The solution of the recursion relations for $\rho$, $\sigma$ and $\mu$ subject to (\ref{boundary}) is  
\be
\rho_n = \lambda - \lambda_2 -n \ ; \qquad \sigma_n = n-\lambda_3 \ ;\qquad \mu_n=\lambda_1-1\ .
\ee
\\
Thus (\ref{ansatz}) reads
\be
\delta_H^\alpha \phi_1=\alpha\sum_{n=0}^{\lambda}(-1)^n \binom{\lambda}{n}\,\partial^{+(\lambda_1 -1)}\left[\bar\partial^n\partial^{+(\lambda -\lambda_2-n)}\phi_2\,\bar\partial^{(\lambda -n)}\partial^{+(n-\lambda_3)}\phi_3\right]+c.c.\: .
\ee
\\
Since
\be
H = \int d^{3}x\,\,\partial_{-}\bar{\phi}_1\,\delta_{\mathcal{H}}\phi_1,
\ee
\\
the interaction Hamiltonian is
\bea
\label{hamilt}
H^{\alpha}=\alpha \int d^{3}x\,\,  \sum_{n=0}^{\lambda}(-1)^n \binom{\lambda}{n}\,\bar{\phi}_1\,\partial^{+\lambda_1}\left[\bar\partial^n\partial^{+(\lambda -\lambda_2 -n)}\phi_2\,\bar\partial^{(\lambda -n)}\partial^{+(n-\lambda_3)}\phi_3\right]+c.c.\: .
\eea
\\
Notice that if we set $\lambda_1=\lambda_2=\lambda_3=\lambda '$ in (\ref{hamilt})with $\lambda '$  odd, $H^\alpha$ vanishes. Hence, a non-vanishing self-interaction Hamiltonian, for odd integer spins exists, if and only if we introduce a gauge group. However, a consistent non-trivial vertex, coupling three fields of different spins, exists irrespective of whether the spins are even or odd.\\ 

\ndt We note that if the action obtained from the above Hamiltonian is to describe a theory involving fields of odd integer spins with cubic self interaction terms, the existence of a gauge group is forced upon the theory. Interestingly, the three fields could, in principle, carry different gauge groups.
 
\vskip 0.3cm

\section{Factorization properties and perturbative relations} 
We now rewrite the above results in the language of spinor helicity~\cite{dixon} where a four-vector is expressed as a bispinor using $p_{a\dot{a}}=p_{\mu}\sigma^{\mu}_{a\dot{a}}$, with $\det(p_{a\dot{a}})$ yielding $-p^{\mu}p_{\mu}$. The spinor products are 
\begin{equation}
<kl>= \sqrt{2}\:\frac{(kl_{-}-lk_{-})}{\sqrt{k_{-}l_{-}}} \qquad  [kl]= \sqrt{2}\:\frac{(\bar{k}l_{-}-\bar{l}k_{-})}{\sqrt{k_{-}l_{-}}} 
\end{equation}
\\ 
Equation (\ref{hamilt}) involves the sum of two kinds of terms: $\!\bar\phi \phi \phi$ and $\phi \bar\phi\bar\phi$. In Fourier space, the coefficient of the second kind of term $\phi_1(p)\bar\phi_2(k)\bar\phi_3(l) \: \delta^{4}(p\!+\!k\!+\!l)$ up to a sign reads
\bea
\frac{p_{-}^{\lambda_1}}{k_{-}^{\lambda_2}l_{-}^{\lambda_3}}
(lk_--l_-k)^{\lambda_2+\lambda_3-\lambda_1},
\eea
\\
which may be rewritten as
\bea
\label{final}
\frac{1}{\sqrt{2^{\lambda}}}\,\,\,<pk>^{^{(-\lambda_1 +\lambda_2 -\lambda_3)}}\:\:<kl>^{^{(\lambda_1 +\lambda_2 +\lambda_3)}}\:\:<lp>^{^{(-\lambda_1 -\lambda_2 +\lambda_3)}}.
\eea
\vspace{.1cm}
\\
It is clear that (\ref{final}) exhibits the nice factorization property described in the introduction. It follows from the expression that given vertices for spins $(\lambda_1, \lambda_2, \lambda_3)$ and $(\lambda '_1, \lambda '_2, \lambda '_3)$, their product yields the vertex for $(\lambda_1 + \lambda '_1, \lambda_2 + \lambda '_2, \lambda_3 + \lambda '_3)$. As a corollary, note that the coefficient for the coupling of three fields $(n\lambda_1 ,n\lambda_2 , n\lambda_3 )$ is the $n$-th power of the coefficient for the coupling $(\lambda_1 ,\lambda_2 , \lambda_3 )$.  
\vskip .4cm
\begin{center}
* ~ * ~ *
\end{center}

\ndt 
This factorization property is similar in spirit to the KLT factorization relations. However, it is important to note that while the KLT relations are on-shell relations concerning amplitudes, the above relations are off-shell and valid at the level of the action. The next logical step in this program of research is to attempt a derivation of consistent quartic interaction vertices for higher spin fields in four dimensional flat spacetime. This step is likely to reveal whether this factorization property exists for higher orders in the interaction. In principle, this would involve the same procedure followed in this paper after incorporating correction terms at order $\alpha^2$. Such a derivation, if successful, would seem to suggest that many no-go results~\cite{BBS}  apply primarily to quantum field theories in which both locality and Lorentz invariance are manifest. Light cone gauge formulations of higher spin theories~\cite{bengt} are very interesting in this regard since locality and Lorentz invariance are no longer manifest and instead need to be checked. The factorization property and the perturbative ties that follow are much broader than the results in~\cite{SA,AA}. They are also very similar in spirit to much of the work devoted to relating spin $1$ and spin $2$ theories~\cite{bohr,bern} but we still have much to learn about higher spin theories perhaps through unitarity methods~\cite{unitary} and S-matrix studies~\cite{smat}.
\vskip 0.5cm
\ndt {\it {Acknowledgments}}
\vskip 0.3cm

\ndt We thank Hidehiko Shimada, Stefano Kovacs and Sunil Mukhi for discussions. We also thank the anonymous referee of~\cite{AA} for suggesting that we generalize our results from cubic self-interactions to cubic interaction of fields with different spins. YSA is supported by an INSPIRE fellowship from the Department of Science and Technology, Government of India.

\end{document}